\documentstyle[graphics,epsfig,floats,aps]{revtex}

\begin{document}

\draft \catcode`\@=11 \catcode`\@=12
\twocolumn[\hsize\textwidth\columnwidth\hsize\csname@twocolumnfalse\endcsname
\title{From Hubbard model to t-J-U model: a canonical transformation
formalism, the metal-insulator transition and mean-field state}
\author{Yue Yu}
\address{Institute of Theoretical Physics, Chinese Academy of
Sciences, P.O. Box 2735, Beijing 100080, China}

\date{\today}

\maketitle

\begin{abstract}
We prove that the t-J-U model can be deduced from the Hubbard
model at a large but finite U by a canonical transformation. We
argue that the system may have a metal-insulator transition at a
critical on-site Coulomb interaction whose value, however, is
smaller than that in previous calculations in which the kinetic
energy has a double counting. In a mean field theory and a special
choice of the parameters, we show that the metallic state may be
equivalent to the gossamer superconducting state proposed by
Laughlin recently.
\end{abstract}

]

The Hubbard model is the generic model for interacting electrons
in the narrow-band and strongly correlated systems \cite{Hubb}.
Especially, since the high temperature superconductivity was
discovered in the cuprates, the Hubbard model on two-dimensional
lattices as well as its strong coupling limit model \cite{andzr},
the t-J model, have been extensively studied  in order to
understand the various anomalous properties of the cuprate
superconductor. The up-date investigations, however, can not
supply a definitive evidence to show the stable $d$-wave
superconducting ground state in these strongly correlated models
\cite{lau}.

It was known that while the Hubbard model is weaker in describing
the exchange correlation, the t-J model is weaker in the long
range charge correlation. A better phenomenological model to
include stronger correlations is the t-J-U model \cite{hua}. The
existence of both J- and U- terms is very important in a possible
mechanism of the high T$_c$ superconductivity, gossamer
superconductivity, proposed by Laughlin recently
\cite{lau,zhang,lau1}. Although including both J- and U- terms in
the model Hamiltonian is physically reasonable for both the
exchange interaction and the on-site Coulomb interaction are
always relevant to the real systems, there is no a microscopic
deduction of the t-J-U model from a generic model.

Comparing with the Hubbard model, there seems to be a double
counting in the t-J-U model to the interacting or kinetic energies
due to taking both the J- and U- terms  into consideration
simultaneously. All previous theories on the t-J-U model did not
deal with this double counting since the lack of a microscopic
understanding. In this paper, we would like to show that the t-J-U
model in fact can be derived from the origin single-band Hubbard
model in a large but finite U. The double counting in the previous
phenomenological t-J-U model was in the kinetic energy as we will
show. An easy understanding  in fact can be seen by checking the
perturbative deduction from the Hubbard model to the t-J model:
The J-term is actually from the second order perturbation in $t/U$
by taking the off-diagonal term for the double-occupied number in
the kinetic term. Namely, if the t-term is decomposed into $T_{\rm
diag}+T_{\rm off-diag}$, the perturbation process transfers
$T_{\rm off-diag}$ to J-term, a virtual hopping process, and only
$T_{\rm diag}$ severs as the real hopping. It can also clearly be
seen from the canonical transformation deduction of the t-J
model\cite{hir}.

In this paper, we would like to deal with the Hubbard model in a
large but finite on-site Coulomb interaction by using the
canonical transformation. We find that the effective Hamiltonian
can be written as the sum over the Hamiltonians acting on a
subspace of the Hilbert space with a fixed double occupied number.
In a partial Gutzwiller projection, we examine the variational
ground of the effective Hamiltonian using the Gutzwiller
approximation. A metal-insulator transition is found at a critical
$U_c$ which however is lower than that estimated by Zhang
\cite{zhang}. The overestimate of $U_c$ in the previous works
stems from the double counting of the kinetic energy in the
phenomenological t-J-U model.

Our model is $U(1)$-invariant due to the particle number
conservation. If we consider a mean field state with the symmetry
breaking, we can show that the mean field state with a special
choice of the parameters is equivalent to Laughlin's gossamer
superconduting state \cite{lau} .

We start from the Hubbard model where the hopping energy may be
dependent on the occupation of sites involved \cite{AAA}.
Including the on-site Coulomb interaction, this Hubbard model
reads
\begin{eqnarray}
H=T+V=T+U\sum_{i=1}^M\nu_i,
\end{eqnarray}
where $M$ is the number of the site;
$\nu_i=n_{i\uparrow}n_{i\downarrow}$,
$n_{i\sigma}=c^\dagger_{i\sigma}c_{i\sigma}$ with $c_{i\sigma}$ a
spin-$\sigma$ electron anahilation operator at site $i$ and the
kinetic term is given by
\begin{eqnarray}
&&T=T_h+T_d+T_{\rm mix},\\
&&T_h=-\sum_{\langle
ij\rangle\sigma}t^h_{ij}(1-n_{i\bar\sigma})c^\dagger_{i\sigma}
c_{j\sigma}(1-n_{j\bar\sigma}),\nonumber\\ &&T_d=-\sum_{\langle
ij\rangle\sigma}t^d_{ij}n_{i\bar\sigma}c^\dagger_{i\sigma}
c_{j\sigma}n_{j\bar\sigma},\nonumber\\
&&T_{\rm mix}=-\sum_{\langle
ij\rangle\sigma}t^{mix}_{ij}n_{i\bar\sigma}c^\dagger_{i\sigma}
c_{j\sigma}(1-n_{j\bar\sigma})\nonumber\\
&&~~~~~~~~~-\sum_{\langle
ij\rangle\sigma}t^{mix}_{ij}(1-n_{i\bar\sigma})c^\dagger_{i\sigma}
c_{j\sigma}n_{j\bar\sigma}.\nonumber
\end{eqnarray}
Except in the mean field theory, we assume
$t^h_{ij}=t^d_{ij}=t^{mix}_{ij}=t$ for the nearest neighbor sites
and vanish otherwise.

In order to define the canonical transformation, we explain our
notations. The partial Gutzwiller projection operator
\begin{eqnarray}
\Pi(g)=\prod_i(1-(1-g)\nu_i) =\sum_{D=0}^{N/2}g^DP_D=g^{\hat D} ,
\end{eqnarray}
where $0\leq g\leq 1$ is the Gutzwiller parameters; $N$ is the
electron number \cite{note}, $\hat D=\sum_i\nu_i$ and
$P_D=\sum_{\{i_1,...,i_D\}}[\nu_{i_1}...\nu_{i_D}\prod_j'(1-\nu_j)]
$ is a projection operator which projects a state into the
subspace with a fixed double-occupation number $D$. $P_0=\Pi(0)$
is the full Gutzwiller projection operator and $\Pi(1)=1$. For
convenience, we denote $ P_D(g)=g^DP_D, P_{\eta_i}(g)=\sum_{D\geq
i}P_D(g)$.

The first goal of this work is to construct an effective
Halmitonian $H_{\rm eff}$ and after the partial Gutzwiller
projection, the projected effective Hamiltonian is given by
\begin{eqnarray}
\Pi(g)H_{\rm eff}\Pi(g)&=&\sum_D P_D(g)H_{\rm
eff}P_D(g)\nonumber\\
&=&\sum_D g^{2D}P_DH_{\rm eff}P_D, \label{ham}
\end{eqnarray}
i.e., all the off-diagonal part $P_{D'}(g)H_{\rm eff}P_D(g)=0$ for
$D'\ne D$. We shall prove that the result effective Hamiltonian in
which all terms keep $D$-invariance is given by
\begin{eqnarray}
H_{\rm eff}=T_h+T_d+{\cal J}+V, \label{Heff}
\end{eqnarray}
where
\begin{eqnarray}
{\cal J}&=&\sum_{\langle ij\rangle} J_{ij}({\bf S_i}\cdot{\bf
S_j}-\frac{1}{4}n_in_j-\frac{1}{2}n_{i\uparrow}n_{i\downarrow}n_j\nonumber\\
&-&\frac{1}{2}n_in_{j\uparrow}n_{j\downarrow}
+n_{i\uparrow}n_{i\downarrow}n_{j\uparrow}n_{j\downarrow} ),
\end{eqnarray}
with $J_{ij}=J\approx 4t^2/U$ for a larger $U$.

The canonical transformation for the Hubbard model to the t-J
model was introduced by Hirsh \cite{hir}. A detailed review for
the canonical transformation can be found in Ref.\cite{bara}. Our
derivation is a generalization of Hirsh's $D=0$ case. Notice that
$P_DT_{\rm mix} P_{D'}=\delta_{D',D\pm 1 }P_DT_{\rm mix}P_{D\pm
1}$ and (\ref{Heff}) remains $D$ invariant, as well as $
\Pi(g)\Pi(g')=\Pi(gg'),~~~P_{\eta_D}(g)P_{\eta_D}(g')
=P_{\eta_D}(gg')$. Keeping these in mind, we do a partial
projection $\Pi(x)H\Pi(x)$ with $x=g^{2/N}$. For large $N$, $x$ is
very close to 1. A straightforward calculation leads to a
rewriting of $\Pi(x)H\Pi(x)$
\begin{eqnarray}
&&\Pi(x)H\Pi(x)=H_0(x)+H_\eta^{(1)}(x),\nonumber\\
&&H_0(x)=H_{diag}(x)+\sum_{D=2}H_\eta^{(D)}(x),
\end{eqnarray}
where
\begin{eqnarray}
&&H_{diag}(x)=\sum_{D=0}P_D(x)HP_D(x),\nonumber\\
&&H^{(D)}_\eta(x)=P_{D-1}(x)TP_D(x)+P_D(x)TP_{D-1}(x).
\end{eqnarray}
The purpose of the canonical transformation is to aquire an
effective Hamiltonian $H_{\rm eff}^{(1)}$ such that $P_0H_{\rm
eff}^{(1)}P_D=P_DH_{\rm eff}^{(1)}P_0=0$ for $D\ne 0$. This
$H_{\rm eff}^{(1)}$ is defined by
\begin{eqnarray}
H_{\rm eff}^{(1)}=e^{i S^{(1)}}\Pi(x)H\Pi(x)e^{-i S^{(1)}}.
\end{eqnarray}
As well-known\cite{hir,bara}, $S^{(1)}$ is determined by the
self-consistent condition $iH_\eta^{(1)}(x)+[H_0(x),S^{(1)}]=0$
and thus the effective Hamiltonian reads
\begin{eqnarray}
H_{\rm
eff}^{(1)}=H_0(x)+\frac{i}2[S^{(1)},H_\eta^{(1)}(x)].\label{h1}
\end{eqnarray}
Solving the self-consistent condition , $H_{\rm eff}^{(1)}$ in a
large U is given by \cite{hir,bara}
\begin{eqnarray} &&P_0H_{\rm eff}^{(1)}P_0\approx
P_0HP_0-\frac{1}{U}P_0HP_{\eta_1}HP_0\nonumber\\
&&~~~~~~~~~~~~~\approx P_0H_{\rm eff}P_0,\label{p0}\\
&&P_{\eta_1}(x)H_{\rm eff}^{(1)}P_{\eta_1}(x)\approx
P_{\eta_1}(x^2)HP_{\eta_1}(x^2).\label{pe}
\end{eqnarray}
From the middle to the right sides in (\ref{p0}), the three site
processes are neglected. If the non-double occupied constraint is
imposed, (\ref{pe}) vanishes because it is related to the double
occupation. Eq.(\ref{p0}) gives rise to the common t-J model.
However, if the double occupation is allowed, we have to deal with
(\ref{pe}). In fact, one can repeats the canonical transformation
to (\ref{pe}) . We would like to require an effective Hamiltonian
$H_{\rm eff}^{(2)}$ whose off-diagonal part $P_1H_{\rm
eff}^{(2)}P_D=P_DH_{\rm eff}^{(2)}P_1=0$ for $D>1$. For this
purpose, one writes
\begin{eqnarray}
\Pi(x)H^{(1)}_{\rm eff}\Pi(x)=P_0H^{(1)}_{\rm eff}P_0+\tilde
H_0(x^2)+H^{(2)}_\eta(x^2),
\end{eqnarray}
where $\tilde H_0(x^2)=P_1(x^2)HP_1(x^2)
+P_{\eta_2}(x^2)HP_{\eta_2}(x^2)$. We do a canonical
transformation and define
\begin{eqnarray}
H_{\rm eff}^{(2)}=e^{iS^{(2)}}\Pi(x)H^{(1)}_{\rm
eff}\Pi(x)e^{-iS^{(2)}},
\end{eqnarray}
where $S^{(2)}$ is required to satisfy $P_0S^{(2)}=S^{(2)}P_0=0$
such that $P_0\Pi(x)H^{(1)}_{\rm eff}\Pi(x)P_0$ is invariant under
the transformation and it is self-consistently determined by
$iH^{(2)}_\eta(x^2)+[\tilde H_0(x^2),S^{(2)}]=0$. Hence, similar
to (\ref{h1}), one has
\begin{eqnarray}
 H_{\rm eff}^{(2)}=P_0H_{\rm
eff}^{(1)}P_0+ \tilde
H_0(x^2)+\frac{i}2[S^{(2)},H_\eta^{(2)}(x^2)].
\end{eqnarray}
Projecting $H_{\rm eff}^{(2)}\to\Pi(x)H_{\rm eff}^{(2)}\Pi(x)$ and
repeating the similar procedure to deduce (\ref{p0}) and
(\ref{pe}), one arrives at
\begin{eqnarray}
&&P_0H_{\rm eff}^{(2)}P_0\approx P_0H_{\rm eff}P_0,\nonumber\\
&&P_1(x)H_{\rm eff}^{(1)}P_1(x)\approx x^2P_1H_{\rm eff}P_1,\nonumber\\
&&P_{\eta_2}(x)H_{\rm eff}^{(2)}P_{\eta_2}(x)\approx
P_{\eta_2}(x^3)HP_{\eta_2}(x^3),
\end{eqnarray}
for a large U, where the three site processes have been ignored.

Repeating this procedure, we finally have
\begin{eqnarray}
\Pi(x)H^{(\frac{N}2)}_{\rm eff}\Pi(x)&=&\sum_{D=0} g^{2D}P_DH_{\rm
eff}P_D\nonumber\\ &=&\Pi(g)H_{\rm eff}\Pi(g). \label{hm}
\end{eqnarray}
The last equality is because $H_{\rm eff}$ is $D$-invariant. The
Gutzwiller parameter is $g$ but not $x$ because we are doing the
partial projection in each time canonical transformation. Thus, we
end the proof of (\ref{ham}) and (\ref{Heff}). Moreover, we see
that, in a partial Gutzwiller projection, the variational ground
state energy is given by a polynomial of the Guztwiller parameter
$g$ in power of $2D$. The coefficient of $g^{2D}$-term is the
ground state energy of the system with a fixed $D$. Using $g$ as a
variational parameter may be convenient for the numerical
simulations. Since a larger U will be considered, the optimal
value of $g$ is expected to be much less than 1. Thus, it may be
enough to calculate only a few terms of the polynomial in the
energy. The numerical work for this model is in progress by using
a variational Monte Carlo simulation \cite{wen}. In this work, we
are going to discuss an infinite system by the Gutzwiller
approximation\cite{gu}. The variational energy per site is given
by
\begin{eqnarray}
\epsilon&=&\frac{\langle \psi_0|\Pi(g) H_{\rm
eff}\Pi(g)|\psi_0\rangle}{M\langle \psi_0|\Pi(g)
\Pi(g)|\psi_0\rangle}\nonumber\\&=& Ud+\langle
(T_h+T_d)\rangle+\langle {\cal J}\rangle.
\end{eqnarray}
Here $d=\frac{\sum_D g^{2D}D/M}{\sum_D g^{2D}}$ is the double
occupied concentrate. Using $d$ as the variational parameter,
Zhang has discussed the metal-insulator transition at a critical
$U_c$ for an effective Hubbard model (or the phenomenological
t-J-U model) \cite{zhang}. As we have mentioned, the Hamiltonian
used in his paper overestimated the kinetic energy, comparing with
our microscopic derivation of the t-J-U model. Meanwhile, we will
see that the critical value of $U_c$ estimated in \cite{zhang} is
higher than that we present here. Taking the approximation
$\sum_{\langle ij\rangle}\nu_in_j\approx 4Dn$ and neglecting the
$\nu_i\nu_j$ term in (\ref{Heff}), the variational condition is
very simple
\begin{eqnarray}
U_{\rm eff}+\frac{\partial g'_t}{\partial d}\langle T\rangle_0+
\frac{\partial g_J}{\partial d}\langle J_{ex}\rangle_0=0.
\end{eqnarray}
here, $J_{ex}$ is the ${\bf S}_i\cdot {\bf S_j}$ term in the
Hamiltonian; $U_{\rm eff}=U-4Jn$ because the contribution from
$\nu_in_j$ term in(\ref{Heff}) and $g'_t$ is modified from the
original Guztwiller factor $g_t$ because of the absence of $T_{\rm
mix}$. $g_J$ is the same as that in the projected Fermi liquid
state \cite{gu}. We have, in fact,
\begin{eqnarray}
&&g_J=\frac{4(1-\delta-2d)^2}{(1+\delta)^2(1-\delta)^2},\nonumber\\
&&g'_t=\frac{2(1-\delta-2d)}{1+\delta}\biggl[\frac{\delta+d}{1-\delta}
+\frac{d(1-\delta)}{(1+\delta)^2}\biggl],
\end{eqnarray}
while $g_t=\frac{8d(1-\delta-2d)}{(1-\delta^2)(1+\delta)^2}$,
where $\delta=1-n$. At half filling, $\delta=0$ and it is easy to
see $g'_t=4d(1-2d)=g_t/2$. Hence, at half filling, the critical
point of the metal-insulator transition is at $U_c=-4\langle
T\rangle_0+16\langle J_{ex}\rangle_0+4J$. Using $\langle
T\rangle_0\approx
 -2\sqrt{2}tC$ and $\langle J_{ex}\rangle_0\approx -(3/4)JC^2$
 with $C\approx 0.479$ \cite{zhang}, $U_c=5.4t+1.25J$. This
 estimate for the critical point is much less than
 $U^{ph}_c=10.8t-2.75J$ which was given by Zhang \cite{zhang}.
 For example, if $J\sim 0.4 t$,
 $U_c\sim 5.9t<U$, while in Ref.\cite{zhang} the
 estimate value $U^{ph}_c\sim 10.1t$. If one estimates $J\approx
 4t^2/U$, $U\sim 10t$ which is near $U_c^{ph}$ but larger than $U_c$.
 This means that for such parameters $J$ and $U$, the
 system is in an absolute insulator state in our theory but is
 near the critical point in the phenomenological model.
 If  $J\sim 0.67t$ ( corresponding to $U\sim 6t$), $U_c\sim 6.2t$ is near U, while in
 Ref.\cite{zhang} the estimate value $U^{ph}_c
 \sim 9.0t>U$. In this case, the system is near the
 critical point in our theory but in an absolute metal state
 in the phenomenological model. Although the
 quantitative value of $U_c$ needs to be refined by numerical
 calculations, the conclusion $U_c<U_c^{ph}$ will not change.

 We now are going to discuss the mean field state of our theory.
 The basic idea to go this mean field state has been explained in
 our previous preprint \cite{yu}, although there were some
 computational errors in that paper. Here we present a renewed version of the
 mean field state.
Introducing two correlation functions $ \Delta_{ij}=\langle
c_{i\downarrow}c_{j\uparrow} - c_{i\uparrow}c_{j\downarrow}
\rangle_0,
 \chi_{ij}=\langle c^\dagger_{i\uparrow}c_{j\uparrow}
+c^\dagger_{i\downarrow}c_{j\downarrow} \rangle_0$, the $U(1)$
symmetry of $H_{\rm eff}$ is broken by a decomposition of the four
particle terms \cite{yu}. According to $\Delta_{ij}$ and
$\chi_{ij}$, the mean field Hamiltonian of (\ref{Heff}) is given
by
\begin{eqnarray}
H_{\rm MF}&=&-\sum_{\langle ij\rangle \sigma}
(t^h_{ij}+t^{(1)}_{ij}(n_{i\bar\sigma}+n_{j\bar\sigma})
+t_{ij}^{(2)}n_{i\bar\sigma}n_{j\bar\sigma})c^\dagger_{i\sigma}
c_{j\sigma} \nonumber \\
&+&\sum_{\langle ij\rangle\sigma}(J_{ij}+J_{ij}^{(1)}(n_{i\sigma}
+n_{j\bar\sigma})+J^{(2)}_{ij}n_{i\sigma}n_{j\bar\sigma})
\nonumber\\ &\times&(-1)^\sigma( \Delta^\dagger_{ij}
c_{i\sigma}c_{j\bar\sigma}+\Delta_{ij}
c^\dagger_{j\bar\sigma}c^\dagger_{i\sigma}) \nonumber\\
&+&U\sum_in_{i\uparrow}n_{i\downarrow}-\sum_{\langle
ij\rangle}2J_{ij}(1-A)n_{i\uparrow}n_{i\downarrow}\\
&-&\sum_{\langle
ij\rangle}J_{ij}(A|\Delta_{ij}|^2+\frac{1}{2}(1-B)|\chi_{ij}|^2)
(n_i+n_j)\nonumber\\ &+&\sum_{\langle
ij\rangle\sigma}J_{ij}(\frac{A}{2}
|\Delta_{ij}|^2n_{i\sigma}n_{j\bar\sigma}+\frac{1-B}{2}|\chi_{ij}|^2
n_{i\sigma} n_{j\sigma}),\nonumber
\end{eqnarray}
where the parameters are given by
\begin{eqnarray}
&&J^{(1)}_{ij}=\frac{A}{2}J_{ij},~~~J_{ij}^{(2)}=-\frac{B}2J_{ij},
\nonumber\\
&&t^{(1)}_{ij}=-t^h_{ij}-(1-A)\chi_{ji}J_{ij},\\
&&t^{(2)}_{ij}=t^h_{ij}+t_{ij}^d-(1-B)\chi_{ji}J_{ij}.\nonumber
\end{eqnarray}
$A$ and $B$ are the variantional parameters to be determined. On
the other hand, we write down Laughlin's gossamer superconducting
Hamiltonian
\begin{eqnarray}
H_G-\mu_RN=\sum_{\bf k}E_{\bf k}\tilde b^\dagger_{{\bf
k}\sigma}\tilde b_{{\bf k}\sigma}, \label{goh}
\end{eqnarray}
where $\mu_R$ is renormalized chemical potential,
$E_k=\sqrt{(\epsilon_k-\mu_R)^2+\Delta_k^2}$ and $\tilde b_{{\bf
k}\sigma}=\Pi(g)b_{{\bf k}\sigma}\Pi^{-1}(g)$ for $ b_{{\bf
k}\uparrow}=u_{\bf k} c_{{\bf k}\uparrow}+v_{\bf k}
c^\dagger_{-{\bf k}\downarrow}$ and $ b_{{\bf k}\downarrow}=u_{\bf
k} c_{{\bf k}\downarrow}-v_{\bf k} c^\dagger_{-{\bf k}\uparrow}$
annihilate the BCS state. Expressing explicitly (\ref{goh}) by the
electron operators \cite{yu,lau1}, we have
\begin{eqnarray}
&&H_G-\mu_R N=-\sum_{\langle ij\rangle
\sigma}[t^G_{ij}+t^{G(1)}_{ij}(n_{i\bar\sigma}+n_{j\bar\sigma})
\nonumber\\&&~~~+t^{G(2)}_{ij}n_{i\bar\sigma}n_{j\bar\sigma}]
+\sum_{\langle ij\rangle \sigma} J_{ij}
[1+\frac{1}2\alpha\beta(n_{i\bar\sigma}+n_{j\bar\sigma})\nonumber\\
&&~~~-\alpha\beta n_{i\bar\sigma}n_{j\bar\sigma}](-1)^\sigma
(\Delta_{ij}^\dagger c_{i\sigma}c_{j\bar\sigma}+\Delta_{ij}
c^\dagger_{j\bar\sigma}c^\dagger_{i\sigma})\nonumber\\
&&~~~+U_G\sum_in_{i\uparrow}n_{i\downarrow}-\mu_G N
\end{eqnarray}
where $\alpha=1-g$ and $\beta=(1-g)/g$ and
\begin{eqnarray}
t^G_{ij}~~&=&t_{ij}^h,\nonumber\\
t^{G(1)}_{ij}&=&-\sum_k\frac{E_{\bf k}}{M}(\alpha v_{\bf
k}^2+\beta u_{\bf k}^2)e^{i{\bf
k}\cdot({\bf r}_i-{\bf r}_j)}\nonumber\\
t_{ij}^{G(2)}&=&\sum_k\frac{E_{\bf k}}{M}(\alpha^2v_{\bf
k}^2-\beta^2u_{\bf k}^2)e^{i{\bf
k}\cdot({\bf r}_i-{\bf r}_j)}, \nonumber\\
J_{ij}\Delta_{ij}&=&\sum_{\bf k}\frac{E_{\bf k}}Mu_{\bf k}v_{\bf
k}e^{i{\bf k}\cdot({\bf r}_i-{\bf r}_j)},
\end{eqnarray}
and $ U_G=\frac{1}{M}\sum_{\bf k}E_k[(2\beta+\beta^2)u_k^2
+(2\alpha-\alpha^2)v^2_k],\mu_G=\frac{1}{M}\sum_{\bf
k}E_k[(2\alpha+1)v_k^2-u_k^2]$. If we identify the t-J-U model to
the gossamer superconducting model in the mean field level, one
requires
\begin{eqnarray}
&& A=\alpha\beta,~~~ B=2\alpha\beta, \nonumber\\ &&
t^{G(1)}=-t^h_{ij}-(1-\alpha\beta)\chi_{ji}J_{ij},\nonumber\\
&&t_{ij}^{G(2)}=t_{ij}^h+t^d_{ij}-(1-2\alpha\beta)\chi_{ji}J_{ij},
\end{eqnarray}
and $\mu_R+\mu_G=J(12A|\Delta_\tau|^2+8(1-B)|\chi_\tau|^2)+\mu,
U=U_G+8J(1-A)$.

Although we have made a formal equivalence between our mean field
state Hamiltonian to Laughlin gossamer superconducting
Hamiltonian, we note that the hopping parameters $t^{G(1,2)}$ have
run out of the practical range in the real materials. Thus, to
show the system described by the t-J-U model has a gossamer
superconducting phase described by Laughlin gossamer
superconducting Hamitonian, a renormalization group analysis is
required. We do not touch this aspect in this work. However, we
can believe there is such a superconducting phase in our theory if
$U< U_c$ because the superconducting paring parameter is
determined by the optimal exchange energy as in the common t-J
model. The renormalization of the hopping parameters is believed
to affect the normal dissipation process only.

In conclusions, we have deduced a t-J-U model from the Hubbard
model at a large but finite U by a series of canonical
transformations. This microscopic derivation showed the
overestimate to the kinetic energy in previous work by a
phenomenological t-J-U model. We then saw that the critical $U_c$
for the metal-insulator transition is much lower than the value
calculated by the phenomenological model. Through a mean field
theory, we showed that the gossamer superconducting state may
possibly be a symmetry breaking state of the t-J-U model as the
BCS common superconducting state versus the Fermi liquid. Another
advantage of our t-J-U model is that the energy was decomposed
into the summation over the energies with the fixed $D$. This is
convenient for numerical simulation.

The author was grateful for the useful discussions to Jingyu Gan,
Jinbin Li, Zhaobin Su, Yuchuan Wen, Tao Xiang, Lu Yu and F. C.
Zhang. This work was supported in part by the NSF of China.


\begin{references}
\bibitem{note} The summation over $D$ runs from 1 to $M$ in
general. However, for a fixed particle number $N$, the highest
number of the double occupied sites is $N/2$.
\end{references}
\end{document}